\documentclass[doublecol]{epl2}
\usepackage{graphicx}%
\usepackage{psfrag}
\usepackage{dcolumn}
\usepackage{amsmath}
\usepackage{amssymb}
\usepackage{amsfonts}
\usepackage{latexsym}
\def\bea{\begin{eqnarray}}
\def\eea{\end{eqnarray}}
\def\beq{\begin{equation}}
\def\eeq{\end{equation}}
\def\f{\frac}

\def\e{\epsilon}

\def\D{\Delta}

\def\t{\tau}
\def\a{\alpha}

\def\rk{\rho^{ (k) }}
\def\rek{\rho^{ (1) }}
\def\cek{C^{ (1) }}
\def\rz{\bar\rho}

\def\r{\rho}

\def\kb{k_B}
\def\la{\langle}
\def\ra{\rangle}
\def\nn{\nonumber}

\def\d{\delta}

\def\l{\lambda}

\def\o{\Omega}

\def\g{\gamma}

\def\a{\alpha}
\def\d{\delta}

\def\la{\langle}
\def\ra{\rangle}
\def\e{\epsilon}
\def\n{\eta}
\def\g{\gamma}

\title{Stochastic pump of interacting particles}
\author{Debasish Chaudhuri\inst{1} \and Abhishek Dhar\inst{2} }
\shortauthor{Debasish Chaudhuri \etal}
\institute{
\inst{1}
FOM Institute AMOLF, Science Park 113, 1098XG Amsterdam, The Netherlands
\thanks{E-mail: \email{d.chaudhuri@amolf.nl}}

\inst{2}
Raman Research Institute, C V Raman Avenue, Sadashivanagar, Bangalore 560080, India
\thanks{E-mail: \email{dabhi@rri.res.in}}
}

\date{\today}

\abstract{
We consider the overdamped motion of Brownian particles, interacting via particle
exclusion, in an external potential that varies with time and space. We show that 
periodic potentials that maintain specific position-dependent phase relations 
generate time-averaged directed current of particles. We obtain analytic results for 
a lattice version of the model using a recently developed perturbative approach. Many 
interesting features like particle-hole symmetry, current reversal with changing 
density, and system-size dependence of current are obtained. We propose possible
experiments to test our predictions.
} 

\begin{document}
\maketitle
\section{Introduction}
Stochastic pumps refer to systems where 
overall directed motion of particles are obtained under the influence of 
external driving forces that are unbiased in the sense that they vanish 
either on spatial or temporal averaging. However, in order to generate 
directed motion, it is necessary that the external forces have some form
of time-reversal-symmetry breaking inbuilt into them. 
Natural examples of stochastic pump include ion-pumps associated with the cell-membrane 
(e.g., ${\rm Na}^+$, ${\rm K}^+$-ATPase pump~\cite{Gadsby2009}), 
molecular motors (kinesin/dynein, myosin) moving on polymeric tracks
(microtubule, F-actin)~\cite{Reimann2002}, etc.  
Ref.~\cite{Astumian2002} showed that the classical pump models for particle
and  heat  transport~\cite{Astumian2001,Reimann2002,Parrondo1998,Marathe2007}  are
similar to isothermal rachet models of molecular motors~\cite{Julicher1997,Reimann2002}. 

Most studies of particle pumps focussed  on systems of non-interacting
particles, though, there have been some numerical~\cite{Derenyi1995} and analytical work~\cite{Derenyi1996} 
on interacting Brownian motors. Recently a model of a classical pump,
similar to those used in the study of quantum pumps~\cite{Brouwer1998,
  Citro2003},  
has been proposed in Ref.~\cite{Marathe2008, Jain2007}. They studied directed
current (DC) in the presence of  inter-particle interactions.
The model studied was the symmetric exclusion process on a ring
which closely mimics a system of diffusing particles with short ranged
repulsive interaction in one dimension. It was shown that for time oscillatory
hopping rates between two neighbouring points a DC current was established
in the system. The hopping rates were taken to be symmetric around every
point, and this made it   
difficult to make a direct connection of this model to a system of particles
in an oscillatory external potentials. In this letter we consider a stochastic
exclusion dynamics which directly mimics hard core particles moving in a 
time-dependent external potential. We show that the perturbative approach 
of Ref.~\cite{Marathe2008, Jain2007} can be used for this case as well, and 
thence obtain a number of interesting predictions for this system.

\section{Model}
The dynamics of small thermally diffusing interacting particles
confined within a narrow tube is well-described by over-damped
Langevin equations in one dimension. Consider $N$ particles interacting with
each other through a short ranged repulsive interaction potential $U$ and in
an external potential $V$. Let $x_r$ denote the position of the $r$-th
particle. The equations of motion are then given by:
$\g {dx_r}/{dt}=-{\partial U}/{\partial x_r}-{\partial V}/{\partial x_r} +
\xi_r~,~ r=1,2,\ldots,N~$, 
where  $\xi_r(t)$ is white Gaussian noise with $\la \xi_r \ra=0$,~
$\la \xi_r(t) \xi_s(t') \ra = 2 \g \kb T \d_{r,s}\d(t-t')$, 
 $\kb$  is the Boltzmann constant and $T$ the ambient temperature. 
We now discuss how this maps to a exclusion process.
In the absence of an external potential, 
hard core colloidal particles of diameter $a$, confined within a narrow channel 
of width $\lesssim 2a$, are well-described by  the symmetric exclusion 
process (SEP)~\cite{Wei2000}. Discretizing space by the particle-size
$a$, the SEP describes
this system with a particle-hopping rate $f_o=D/a^2$ where the diffusion constant
$D$ obeys the Einstein relation $D=\kb T/\g$. 
The effect of an external potential $V(x,t)$
is to make the hopping rates time- and position- dependent. 
Let $P(x+a,t+\delta t|x,t)$ be the conditional probability density of a
particle being at $x+a$ at time $t+\delta t$, given that it is at $x$ at time
$t$. Then from the Langevin
dynamics we see that the transition rate is given by 
$P(x+a, t+\d t \mid x, t)~a~/\d t \sim f_o [1-(V(x+a)-V(x))/2 \kb T]$,
for weak potential differences [$(V(x+a)-V(x))< 2\kb T$].

Thus we consider  a lattice model consisting of $N$ particles on a ring of $L$
sites with lattice spacing $a$.  Each site $l=1,2,\dots,L$  may contain $n_l=0,1$ particle. 
A particle at site $l$ can hop to an empty nearest neighbour  site $l \pm 1$
with rate $w_{l,l\pm 1}=f_0[1-\l(u_{l\pm 1}-u_l)/2]$, where $\l u_l = V_l/\kb T$. 
We consider driving at a constant frequency  and set 
$u_l =  \a_l \sin(\o t +\phi_l)=2\,\mbox{Re}~ [\n_l e^{i \o t}] $
where $\n_l =\a_l  e^{i \phi_l}/2i$.
For time-independent potential $V_l$, i.e., $\o=0$, the transition rates satisfy detailed
balance and the system reaches thermal equilibrium with zero current. As we show now,
in the presence of time-dependent $V_l$, the system is driven out of equilibrium and can show 
directed flow of particles resulting from AC driving forces.  We analyze the model utilizing a 
perturbative expansion in the strength $\l$ of time-periodic potential, and also via Monte-Carlo simulations.

\section{Calculation of the current}
The current from site $l-1$ to $l$,  $J_{l-1,l}$ is defined by the continuity
equation obtained from the time evolution of local density $\r_l = \la n_l
\ra$, where $\la...\ra$ denotes an average over the noise. 
This gives~\cite{Schutz2000} 
$d\r_l/dt = J_{l-1,l} - J_{l,l+1}$
where
\beq
J_{l-1,l} = (w_{l-1,l}\,\r_{l-1} - w_{l,l-1}\, \r_l) - (w_{l-1,l}-w_{l,l-1})\,C_{l-1,l} 
\eeq
gives the noise averaged current from site $l-1$ to site $l$ and
 $C_{l,m} = \la n_l n_m \ra$ denotes the two-point correlations.
$C_{l,m}$ appears in the expression of $J_{l-1,l}$ due to the particle exclusion.
In the time-periodic steady state the current on each bond, averaged over the time period 
$\t = 2\pi/\o$, is the same and hence the net DC current is given by 
$\bar J = (1/L\t)\sum_{l=1}^L \int_0^\t dt J_{l-1,l}(t)$.
Using the form of the hopping rates we then get:
\bea
\bar J 
= -\f{\l f_0}{2L\t}\sum_{l=1}^L \int_0^\t dt\, (u_l-u_{l-1})(\r_{l-1}+\r_l - 2 C_{l-1,l}).
\label{eq:jbar}
\eea
To evaluate $\bar J$ we need to evaluate $\r_l$s and $C_{l-1,l}$s and this we 
do using perturbation theory.  For weak driving potential we make a perturbative 
expansion of $\r_l$, $C_{l,m}$ in a series in the dimensionless parameter $\lambda$
\bea
\r_l &=& \rz + \sum_{k=1,2,\dots} \l^k \rk_l, \crcr
C_{l,m} &=& C_{l,m}^{(0)} + \sum_{k=1,2,\dots} \l^k C_{l,m}^{(k)}.
\eea
The $\l = 0$ terms in the above expansions correspond to the absence of any
external potential and this is then 
just the symmetric  exclusion process  with uniform hopping rate $f_0$. For that case the steady state
is the equilibrium state obeying detailed balance and is characterized by the position-independent densities and
correlations. These are all known exactly, for example
$\rz = {N}/{L}=\r $, $C_{l,m}^{(0)}  =  \r (N-1)/(L-1) $, 
etc.~\cite{Schutz2000}.

As noted in Ref.~\cite{Marathe2008} 
the time evolution of the first order terms $\rek_l(t)$ and $\cek_{l,m}(t)$  are given by the
following equations:
\begin{widetext}
\bea
\f{d\rek_l}{dt} &=& f_0 \D_l \rek_l + f_0 q_0 \D_l u_l \, ,   
\label{eq:rek} \\
\f{d \cek_{l,m}}{dt} &=& f_0 (\D_l + \D_m)\cek_{l,m} + f_0 k_0 (\D_l u_l+ \D_m u_m) ~~ {\rm for} ~ l\neq m \pm 1, \crcr
\f{d \cek_{l,l+1}}{dt} &=& f_0 (\cek_{l-1,l+1} + \cek_{l,l+2} -2 \cek_{l,l+1} )+ f_0 k_0 (u_{l-1}+u_{l+2}-u_l-u_{l+1})
\label{eq:cek}
\eea
\end{widetext}
\begin{floatequation} 
\mbox{\textit{see eq.~\eqref{eq:rek},~\eqref{eq:cek}}} 
\end{floatequation}
where $\D_l g_{l,m} = g_{l+1,m} + g_{l-1,m} - 2 g_{l,m}$ defines the discrete Laplacian, and 
$q_0=\rz-C_{l,m}^{(0)}$, $k_0 = C_{l,m}^{(0)} - C_{l,m,n}^{(0)}$ with 
{\em equilibrium} three-point correlation $C_{l,m,n}^{(0)}=C_{l,m}^{(0)} (N-2)/(L-2)$.
These linear  homogeneous equations can be solved exactly to obtain 
the long-time oscillatory  solution~\cite{Marathe2008}. We find:  
\bea
\rek_l(t) = 2 {\rm Re} [A_l e^{i \o t}]
\eea
where the vector ${\vect A} = \{A_1, A_2,\dots,A_L\}$ is given by 
\beq
{\vect A} = \f{q_0 f_0}{i\o -f_0 \hat \D} \hat \D {\vect \n} =-q_0 {\vect \n} + \f{i \o q_0}{i\o -f_0 \hat \D} {\vect \n}
\eeq
with $\vect{\n}=\{\n_1,\n_2,\dots,\n_L\}$ and $\hat \D$ the discrete Laplacian operator.
The eigenfunctions of $\hat \D$ are given by $(1/\sqrt L) e^{-i q l}$ with
corresponding eigenvalues $\e_q=-2(1-\cos q)$. 
Expanding $\hat \D$ using  its eigenbasis we get $A_l = -q_0 \n_l + \sum_m R_{l,m} \n_m$   
where $R_{l,m}=(i\o q_0/L) \sum_q e^{i q (l-m)}/(i\o - f_0 \e_q)$ with $q=2\pi
j/L$, $j=1,2,\dots,L$. 
The equation of two-point correlations can also be solved and it is easy to
verify the following  
steady state solutions (for all $l,m$):
\beq
\cek_{l,m} (t) = \f{k_0}{q_0} [ \rek_l(t) + \rek_m(t)].
\label{eq:corr}
\eeq

Then using Eq.~\ref{eq:jbar} one can find the general expression 
for the directed current,
\bea
\bar J &=& -\f{\l^2 f_0}{2L} \left(1-\f{2 k_0}{q_0} \right) 
\sum_{l=1}^L 2\,{\rm Re} \left[ (\eta_l^\ast - \eta_{l-1}^\ast) (A_{l-1} +
  A_l) \right] \nn \\  
&=& -\f{\l^2 f_0}{L} \left(1-\f{2 k_0}{q_0} \right) {\rm Re}\left[ \sum_{l,m} \n_l^\ast \left(R_{l-1,m} - R_{l+1,m}\right) \n_m \right]\nn  \\
&=& -\f{2 \l^2 f_0 \o}{L} (q_0 -2 k_0) {\rm Re} \left[ \sum_q \mid \tilde \n_q \mid^2 \f{\sin q}{i\o - f_0 \e_q}\right]
\label{eq:J}
\eea
where $\tilde \n_q = (1/\sqrt L) \sum_l \n_l e^{-iq l}$.

Note that the effect of interaction is entirely contained in the prefactor $(q_0-2k_0)$, which in the large $L$ limit 
equals to $\r(1-\r)(1-2\r)$. Several interesting features follow. The current vanishes at half-filling and its sign reverses
with increasing particle density. The dynamics has particle-hole symmetry, and this is explicit in the density dependence.
At low densities we obtain $\bar J\propto \r$ and this corresponds to the case where interactions can be neglected.
This differs from the result in Ref.~\cite{Marathe2008} where the current vanished in the absence of interactions.
We now investigate the solution in Eq.~\ref{eq:J} for different choices of the oscillating potential.
We consider the following two cases:

\section{Case (i) : Localized pump}
Here we consider the specific case of a localized pump with time-varying potentials acting only on two consecutive sites
such that $\a_1=\a_L=1$ with all other $\a_l=0$, and %$\phi_1=0$, $\phi_L=-\phi$.
$\phi_1=\phi$, $\phi_L=0$.
For this case Eq.~\ref{eq:J} leads to the DC current 
\bea
\bar J_2 = -\l^2 (q_0-2 k_0)  \f{\o \sin\phi}{2L}\mbox{Re}\left[\f{z_- - z_-^{L-1}}{1-z_-^L}\right],  
\label{eq:J2}
\eea
where $z_- = y/2 - \sqrt{(y/2)^2-1}$ with $y=2+i\o/f_0$.
Note that $\bar J_2$ has a sinusoidal dependence on the phase difference of external driving $\phi$, leading to
maximal driven current at $\phi=\pm \pi/2$. $\bar J_2$ decays as $1/L$ with the system size $L$.

\begin{figure}[t] % [htbp]
\begin{center}
\psfrag{J}{$\bar J_2\, \times\, 10^4$}
\psfrag{rho}{$\r$}
\includegraphics[width=7 cm] {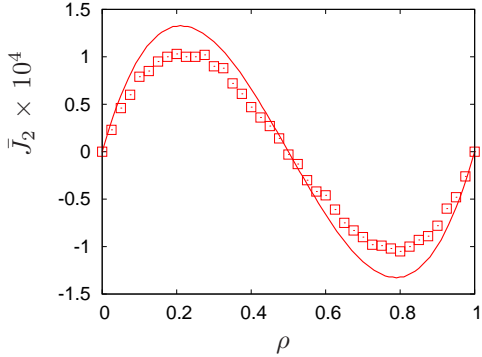}  
\caption{(Color online) 
Directed current $\bar J_2$ as a function of mean density  $\r$. The points denote Monte-Carlo results and the line is a plot 
of the function in  Eq.\ref{eq:J2}. The parameters used are: system size
$L=16$, hopping rate $f_0=0.34$,  potential strength $\l=0.5$, time period of
oscillation $\t=10$ ($\o=0.2\pi$), and phase difference $\phi=-\pi/2$.  
The data (points) were collected over $100\t$ after equilibration over another $100\t$. All the data were averaged over $10^6$ initial conditions.}
\label{fig:Jro}
\end{center}
\end{figure}
In Fig.~\ref{fig:Jro} we plot density dependence of $\bar J_2$   obtained from Monte-Carlo simulations.
The data show good agreement with Eq.~\ref{eq:J2}.
At $\r=1$ the current vanishes due to complete jamming of particles. Note the zero in 
the current at the point of half-filling $\r=1/2$, and the reversal in the direction of current as the density crosses this point (Fig.~\ref{fig:Jro}). 
The maximal currents are obtained at $\r = (1 \pm 1/\sqrt 3)/2$. We re-emphasize that the density 
dependence is general, and independent of whether the pumping is localized or 
spatially distributed.

\section{Case (ii) : Traveling wave}
\begin{figure}[t] % [htbp]
\begin{center}
\psfrag{J}{$\bar J\, \times\, 10^4$}
\psfrag{omg}{$\o$}
\psfrag{J      }{$\bar J_2 L$}
\psfrag{JL      }{$\bar J_L$} 
\includegraphics[width=7 cm] {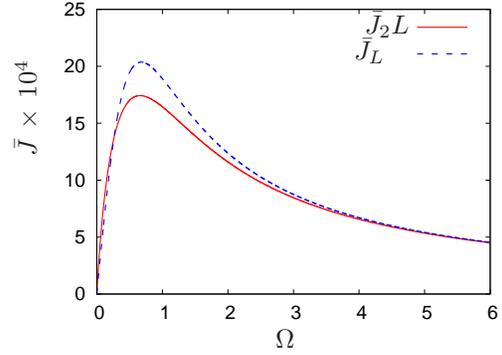}  
\caption{(Color online) 
System-size independent currents $\bar J_2 L$ and $\bar J_L$ as a function of
frequency $\o$ using Eqs.~\ref{eq:J2} and \ref{eq:JL} with
$L=10^4$ and $\r = (1-1/\sqrt 3)/2$. All other parameters are the same as in Fig.~\ref{fig:Jro}.
The maximum in $\bar J_L$ is at $\o=2 f_0 = 0.68$.
}
\label{fig:Jomega}
\end{center}
\end{figure}
A spatially distributed external potential of the form  $V(x,t)=V_0 \sin(\o t + \phi x)$ 
where $\phi$ is constant generates a directed traveling wave of external force. 
We use $\a_l =1$ and $\phi_l=\phi\, l$ with a constant $\phi$,
i.e., $\n_l = \exp(i \phi l)/2i$ for  all $l$. 
Hence we have $\tilde \n_q = (1/2i) \sqrt L \d_{q,\phi}$, 
and using this expression in  Eq.~\ref{eq:J}  we find the time-averaged
current   
\bea
\bar J_L &=& -\f{\l^2 f_0 \o}{2} (q_0 -2 k_0) {\rm Re} \left[\f{\sin \phi}{i\o - f_0 \e_\phi}\right] \crcr
&=&  -\l^2 (q_0 - 2 k_0) f_0^2  \f{\o  \sin\phi(1-\cos\phi)}{\o^2+4 f_0^2(1-\cos\phi)^2}.
\label{eq:JL}
\eea
Since in this case external force acts on all the sites of the system  $\bar J_L$  is  independent  of  system size $L$. This behaviour is in contrast to the decay of current in localized pump as $\bar J_2 \propto 1/L$.
While Eqs.~\ref{eq:J2} and \ref{eq:JL} contain the full functional dependence of the
time-averaged currents for the two cases, it is instructive to illustrate their frequency-dependence  
at a density and phase at which the currents are at their maximum. 
Thus, using  $\phi=-\pi/2$ for a system of density $\r=(1-1/\sqrt 3)/2$ and a large system size $L=10^4$, 
the dependence of $\bar J_2 L,\, \bar J_L$ on driving frequency $\o$ is shown in Fig.~\ref{fig:Jomega}.
At small frequencies, current grows linearly with $\o$. 
Eq.~\ref{eq:JL} shows that $\bar J_L$ reaches a maximum value $\bar J_L^{\rm max} = \l^2 f_0/24\sqrt 3$ at a driving frequency $\o=2 f_0$.  
This denotes a resonance between the intrinsic relaxation rate $f_0$ and the 
external driving frequency $\o$. At high frequencies $\o> 2f_0$,
the relaxation of individual particles lags behind the fast variation of external force, 
and eventually the current decays as $1/\o$.

\section{Outlook}
Our model of stochastic exclusion dynamics 
showed emergence of DC current in response to periodic forcing that vanishes 
on temporal averaging. 
Definite phase-difference between the oscillatory forcing at neighbouring sites
breaks the time-reversal symmetry to yield directed current.
The mechanism works even in the limit of very low densities where particles rarely 
interact. However, with increasing density we found a non-monotonic variation of the DC current, entirely due to the exclusion interaction between particles. Most interestingly, 
the system showed a current reversal around half filling: At density $>1/2$, the driven DC current flows 
in a direction opposite to that in the non-interacting limit. This behavior is  associated with the particle-hole symmetry --
the dynamics of holes at high densities is equivalent to the dynamics of particles at low densities.
 
Our predictions may be tested experimentally by using oscillatory force on
colloids moving in confined geometries. 
There have been several experimental observations in colloidal systems of
single-file-diffusion which is  one of the signatures of diffusing interacting
particles moving in one dimension
\cite{Hahn1996,Kukla1996,Wei2000,Lutz2004PRL,Lutz2004}. It is possible to
apply localized AC drive on such confined-colloids by, e.g., exploiting AC
electro-kinetic effects, or localized oscillatory pressure.    
Alternatively in the setup of Ref.\cite{Lutz2004PRL} one could impose
oscillatory trapping potential by  using laser traps of tunable power.   
Consider the motion of colloidal particles of diameter % radius $a \approx 2 \mu $m 
$a \approx 2 \mu $m  moving
in water inside  a narrow tube of diameter $ \lesssim 2 a$. 
At room temperature $\kb T = 4.2 \times 10^{-21}\,$N-m using water viscosity
$\nu \approx  10^{-3}$ ${\rm N s/m}^2$ one finds the Stokes-Einstein
self-diffusion constant  
$D_0 = \kb T/3\pi\nu a \approx 0.2\, \mu {\rm m}^2/$s. 
This gives hopping rate %$f_o =D_0/a^2 =0.025\, {\rm s}^{-1}$. 
$f_o =D_0/a^2 =0.05\, {\rm s}^{-1}$. 
Then for the case of extended
driving corresponding to the result in Eq.~\ref{eq:JL}, the maximum current is
obtained at a density of $\rho=(1-1/\sqrt 3)/2 \approx 0.2$ and a driving frequency of  
$\Omega=2 f_o\approx 0.1\,$Hz. Since the
value of the maximal current $\bar{J}_L \approx \lambda^2 f_o/40$, this means that
we can get particle flow velocities $\sim \bar J_L a \sim 0.0025\, \mu {\rm m/s} $. 
This compares with the drift velocity that would be attained by a
colloidal particle of the same radius and carrying a charge of $10$  electrons, when placed in an electric field 
$\sim 30\,$V/m.

\acknowledgements
The work of DC is part of the research program of the ``Stichting voor Fundamenteel 
Onderzoek der Materie (FOM)", which is financially supported by the ``Nederlandse 
organisatie voor Wetenschappelijk Onderzoek (NWO)". DC thanks Raman Research 
Institute for hospitality during a short visit which initiated this work,
and Nils Becker for a critical reading of the manuscript.

\bibliographystyle{prsty}

\end{document}